\documentclass[sigconf]{acmart}
\AtBeginDocument{%
  \providecommand\BibTeX{{%
    \normalfont B\kern-0.5em{\scshape i\kern-0.25em b}\kern-0.8em\TeX}}}

\usepackage{url}
\usepackage{cleveref}
\usepackage{multicol}
\usepackage{multirow}
\usepackage{booktabs}
\usepackage{enumitem}
\usepackage{comment}
\usepackage{siunitx}

\usepackage{xspace}
\usepackage{booktabs}

\title{What Types of Questions Require Conversation to Answer? A Case Study of AskReddit Questions}

\definecolor{ao(english)}{rgb}{0.0, 0.5, 0.0}

\newcommand{\kenneth}[1]{}
\newcommand{\alan}[1]{}
\newcommand{\cy}[1]{}
\newcommand{\avon}[1]{}
\newcommand{\hua}[1]{}
\newcommand{\ed}[1]{}
\newcommand{\ryan}[1]{}
\newcommand{\gromit}[1]{}

\newcommand{\convsys}{conversational systems\xspace}
    
\newcommand{\eg}{{\it e.g.}}
\newcommand{\ie}{{\it i.e.}}

\newcommand{\qone}{Reach-Out\xspace}
\newcommand{\qtwo}{Scope\xspace}
\newcommand{\qthree}{Eliciting\xspace}
\newcommand{\qfour}{Elaboration\xspace}
\newcommand{\qfive}{Duration\xspace}
\newcommand{\qsix}{Conversation\xspace}
\newcommand{\qseven}{Format\xspace}
\newcommand{\qeight}{Difficulty\xspace}

\copyrightyear{2023}
\acmYear{2023}
\setcopyright{rightsretained}
\acmConference[CHI EA '23]{Extended Abstracts of the 2023 CHI Conference on Human Factors in Computing Systems}{April 23--28, 2023}{Hamburg, Germany}
\acmBooktitle{Extended Abstracts of the 2023 CHI Conference on Human Factors in Computing Systems (CHI EA '23), April 23--28, 2023, Hamburg, Germany}\acmDOI{10.1145/3544549.3585600}
\acmISBN{978-1-4503-9422-2/23/04}
\begin{document}


\author{Shih-Hong Huang}
\affiliation{%
  \institution{The Pennsylvania State University}
  \city{University Park}
  \state{PA}
  \country{USA}
}
\email{szh277@psu.edu}

\author{Chieh-Yang Huang}
\affiliation{%
  \institution{The Pennsylvania State University}
  \city{University Park}
  \state{PA}
  \country{USA}
}
\email{chiehyang@psu.edu}

\author{Ya-Fang Lin}
\affiliation{%
  \institution{The Pennsylvania State University}
  \city{University Park}
  \state{PA}
  \country{USA}
}
\email{yml5563@psu.edu}

\author{Ting-Hao `Kenneth' Huang}
\affiliation{%
  \institution{The Pennsylvania State University}
  \city{University Park}
  \state{PA}
  \country{USA}
}
\email{txh710@psu.edu}

\renewcommand{\shortauthors}{Huang, et al.}

\begin{abstract}
The proliferation of automated conversational systems such as chatbots, spoken-dialogue systems, and smart speakers, has significantly impacted modern digital life. 
However, these systems are primarily designed to provide answers to well-defined questions rather than to support users in exploring complex, ill-defined questions.
In this paper, we aim to push the boundaries of conversational systems by examining the types of nebulous, open-ended questions that can best be answered through conversation. 
We first sampled 500 questions from one million open-ended requests posted on AskReddit, and then recruited online crowd workers to answer eight inquiries about these questions. 
We also performed open coding to categorize the questions into 27 different domains. 
We found that the issues people believe require conversation to resolve satisfactorily are highly social and personal.
Our work provides insights into how future research could be geared to align with users' needs.

\end{abstract}





\maketitle

\section{Introduction}

Automated conversational systems such as chatbots, spoken-dialogue systems, and smart speakers have become routine in modern digital life.
With recent advances in deep learning, today's cutting-edge conversational systems can produce fluent responses to users' messages, find pieces of information as requested, and execute simple voice commands.
These systems are designed to quickly deliver concrete answers to well-defined questions.
However, the potential of human-to-AI conversations extends beyond this. 
People have been solving difficult issues by talking to each other for thousands of years.
Interaction allows conversational partners to explore ill-defined, complicated problems together. Open-ended discussion allows people to shape their thoughts and stances on complex issues. 
Unfortunately, the literature has little to say about how conversational systems can be built to support, facilitate, or even participate in such important discussions.
Most task-oriented conversational systems have been built with a relatively clear task procedure in mind, \eg, typical user intents, what information is needed to fulfill each intent, steps to take to elicit needed information from the user, and how to accomplish a task.
But real-world problems are usually imprecise. The structure, procedure, needed information, and end goals are often unclear or undecided.
Everyday questions as common as ``What kind of dog should I get?'' and ``How can I fit into a new environment?"
often require back-and-forth discussion to form a helpful answer and can be vastly different for different people.
Although chatbots powered by language models such as ChatGPT~\cite{chatgpt} and YouChat~\cite{youchat} can engage in open-ended conversations to some extent, they are not primarily designed to solve complicated real-world tasks. Instead, they focus on generating human-like responses to various prompts and inputs.

\begin{figure*}
    \centering
    \includegraphics[width=0.99\linewidth]{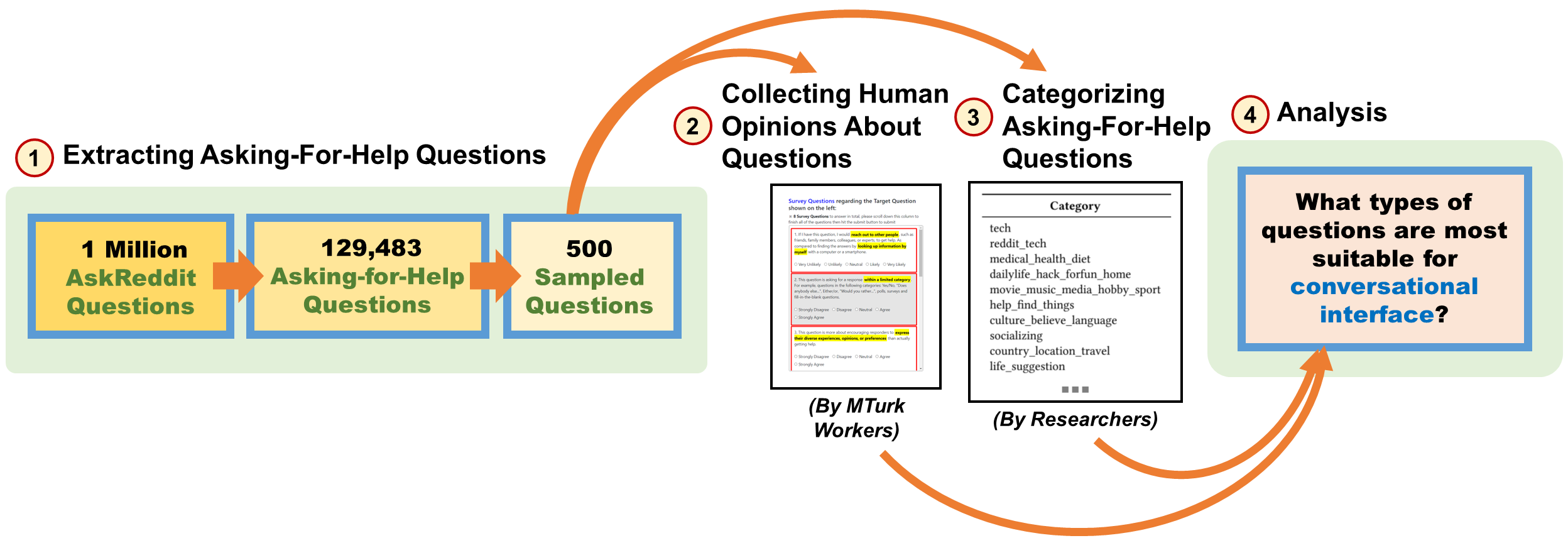}
    \vspace{-.5pc}
    \caption{The study procedure overview. We (1) sampled 500 asking-for-help questions from one million random AskReddit questions and     (2) recruited crowd workers to answer eight inquiries about these questions.
        Furthermore, we (3) performed open coding on the asking-for-help questions to categorize them, allowing us to (4) gain insight into which sorts of topics require conversation the most.}
    \Description{Procedure overview of the study separated into four blocks from left pointing to the right, labeled in 1. Extracting Asking-for-Help Questions, 2. Collecting Human Perceptions Toward Questions, 3. Categorize Asking-For-Help Questions, and 4. Analysis. Block number one contains three small blocks pointing from left to right. The three small blocks are, 1 Million AskReddit Questions, 129,483 Asking-for-Help Questions, and 500 Sampled Questions. Block 1 points to block 2 and 3, respectively. Block 2 shows the MTurk interface for workers to demonstrate collecting human opinions. Block 3 represents the labeling process by the researchers. Both block 2 and 3 point to block 4. The text in block 4 is "What types of questions are most suitable for conversational interface?".}
    \label{fig:procedure}
\end{figure*}

In this paper, we aim to push the boundaries of conversational systems by examining the types of ill-defined, open-ended questions that can best be answered through conversation.
We studied the questions people posted on r/AskReddit\footnote{https://www.reddit.com/r/AskReddit/}, which tend to be open-ended and loosely defined.
AskReddit is an online discussion board (subreddit) of Reddit, a platform on which users can submit open-ended questions to which other users then respond.
We extracted one million random questions from the AskReddit subreddit and created a machine-learning classifier to identify questions that asked for help; it identified 129,483.
Then we recruited online crowd workers from Amazon Mechanical Turk to answer eight inquiries about 500 randomly sampled asking-for-help questions.
These inquiries indicated how much the question required a conversation to be satisfactorily resolved and how the user would most want to get it answered.
For example, in one of the inquiries (Q7), 
we told workers to assume they were asking the question using a computer or smartphone and asked how much they would prefer the answer be provided through a conversation as compared to written formats such as emails.
Finally, we performed open coding on the asking-for-help questions to categorize them into 27 different domains, allowing us to analyze which topics MTurk workers believed would require conversation the most. \Cref{fig:procedure} shows the procedure overview of the study.

Instead of asking what a conversational system can do, this work takes a step back and uses a data-driven approach to ask what people \textit{hope} conversational systems can do for them.
Our work will inform the development of future systems and help us reflect on the current status of chatbots, spoken dialogue systems, and smart speakers.

\section{Background}

\kenneth{Cite more paper? This part will need some revision.}
Being able to hold human-like open-domain conversations is one of the biggest challenges in AI.
With recent advances in large language models, today's cutting-edge conversational systems are capable of producing fluent responses to users' messages~\cite{51348} and reliably finding requested information~\cite{zamani2022conversational,gao2022neural}.
The true value of human conversation lies beyond lightweight chitchat or solving clearly defined tasks such as booking a flight.
People talk to each other to navigate complex, ill-defined problems together.
Modern intelligent assistants such as Amazon's Echo promise a future in which conversing with a machine is as easy as talking to a friend.
But these conversational systems' capacity is still far from what talking to a friend can offer. While the latest language models like ChatGPT~\cite{chatgpt} and YouChat~\cite{youchat} are capable of interacting with the users in a conversational manner, concerns regarding the correctness of the responses provided by such models have been raised, and the limitations of such models are also unclear~\cite{https://doi.org/10.48550/arxiv.2212.05856}.\avon{any citations showing the concerns?}\cy{Probably cite this one? \url{https://arxiv.org/abs/2212.05856}. I don't think we have enough formal publications discussing chatGPT now as it was released only a month ago.}

Researchers have attempted to bootstrap open-domain conversational systems. A classic example is Evorus, which was initially a human-powered chatbot operated by online crowd workers~\cite{chorusDeploy,ignition2017} that automated itself over time~\cite{huang2018chi}.
Evorus had crowd workers use a worker interface to propose responses, take notes, and vote to sort others' replies and identify optimal responses.
These collective actions allowed the crowd to converse with the user as a single, consistent conversational partner.
More importantly, each action the workers took could be automated over time to gradually move away from human-powered systems.
However, one lesson learned from Evorus was that more research is needed to create conversational systems that can solve ill-defined problems~\cite{huang2019automating}.
Real-world problems are often complicated and imprecise, and a universally optimal solution may not exist.
Supporting or automating such conversations requires approaches beyond taking notes and sorting ideas.







\section{Methods}
\begin{table*}
    \centering \small
    \begin{tabular}{lp{11cm}}
        \toprule
        \multicolumn{1}{c}{\textbf{Label}} & \multicolumn{1}{c}{\textbf{Questions}} \\ \midrule
        Asking-for-help                       &What tasks can only be accomplished by humans, and cannot be accomplished by AI or robots?  \\ \midrule
        Asking-for-help                       &How to increase reddit trophies and how to get it easily ?  \\ \midrule
        Asking-for-help                       &If a dog scratches you and doesn’t bleed but leaves a mark, will it scar?  \\ \midrule
        Asking-for-help                       &Trying to get my drivers license after having my permit for 6 months what do i do ?  \\ \midrule
        Asking-for-help                       &Vietnam war has ``fortunate son'' as its theme song. What other war has a theme song? \\ \midrule
        Others                                &If you were a computer what would your specs be? \\ \midrule
        Others                                &What was your favourite period in your life?  \\ \midrule
        Others                                &If you had to choose a famous person to swap lives with, who would it be?  \\ \midrule
        Others                                &People of reddit who taught themselves in anything how and why did you do it?  \\ \midrule
        Others                                &What is life like for you now?  \\
        \bottomrule
        \end{tabular}
    \caption{Examples of ``Asking-for-help'' and ``Others'' categories of questions. Asking-for-help questions are defined as questions people will ask a single agent that have a finite answer.}
    \label{tab:ask_for_help_example}
\end{table*}
\begin{table}
    \centering \small
    \begin{tabular}{cccc}
    \toprule
	    & \textbf{Asking-for-help} & \textbf{Others} & \textbf{Total} \\ \midrule
    \textbf{All}   & 133 & 1,859 & 1,992 \\
    \textbf{Train} & 111 & 1,483 & 1,594 \\
    \textbf{Valid} & 22  & 376  & 398  \\ 
    \bottomrule
    \end{tabular}
    \caption{Data statistics of the 1,992 annotated Asking-for-help Reddit dataset questions. We split data into train and valid sets using a ratio of 0.8 and 0.2 respectively.}
    \label{tab:data-statistic}
\end{table}

\subsection{Data Preparation}
We extracted questions from the \texttt{one-million-reddit-questions} dataset~\cite{reddit-qa}.
The million questions covered a variety of topics and included questions such as ``What is the best story in your family?'', ``What frustrates you more than anything?'', ``What language/s do you speak?'', and so on.
We noticed that the data contained many questions that were meant to engage a large audience on Reddit to elicit responses with diverse viewpoints rather than asking for help. (\Cref{tab:ask_for_help_example} shows some examples of asking-for-help questions.)\avon{x}\alan{rather than to solve a problem}\avon{I feel like "ask for help" could have various meanings. e.g. providing personal preference could be a kind of help? How about "problem-solving" (I am not sure though...)}
In this paper, we focused on questions that can benefit from one-on-one conversations with a single conversational partner rather than with a crowd.
Therefore, we first built a classifier to extract questions that were actually \textbf{asking for help}.

\subsubsection{Building a Classifier to Extract Asking-for-Help\avon{discussed with Alan, should change to "Asking-for-help"}\avon{change to "Ask-for-help"?} Questions}
To train the classifier to identify questions that were asking for help, one of the authors (A1) annotated 1,992 randomly sampled questions from the dataset of one million Reddit questions (L1). 
As we are interested in those questions that can be responded to by a conversational agent, questions that people would ask a single ideal agent and have a finite answer were considered valid.
\alan{As we are interested in those questions that can be solved by a conversational agent, questions that people will ask a single ideal agent and should have a finite answer were considered valid.}%
\avon{Why it should have finite answer to be an ask-for-help question? How about "As we are interested in questions conversational agent could solve, questions that were considered as valid were those that people will ask a single ideal agent and should have a finite answer."}%
Questions that were not considered as asking-for-help questions were those based on personal opinions and experiences of the answerer. 
Questions intended to generate debate and instigate conflict were also excluded. 
Inter-coder reliability was investigated by another author (A2) independently annotating 450 randomly sampled questions out of the 1,992 questions (L2) following the coding criteria. 
The research team also discussed and generated another set of annotations (L3) for the 450 randomly sampled questions that were agreed upon among A1, A2, and another research team member (A3). 
The inter-coder reliability (Cohen's kappa $\kappa$) between each of the annotations (L1, L2, L3) was $\kappa(L1, L2) = 0.545$, $\kappa(L1, L3) = 0.748$, and $\kappa(L2, L3) = 0.802$.
\Cref{tab:ask_for_help_example} shows some example sentences;
\Cref{tab:data-statistic} shows the statistics indicating data imbalance.
The data was split into train/valid sets using the ratio 0.8 and 0.2 respectively.
We then fine-tuned DeBERTa~\cite{he2021deberta} (\texttt{microsoft/deberta-v2-xxlarge}) using Pytorch~\cite{NEURIPS2019_9015} and Huggingface~\cite{wolf-etal-2020-transformers} for text classification.
The hyperparameters used were batch size = 32, learning rate = 1e-5 with the linear scheduler, and warm-up ratio = 0.05.
The model was fine-tuned with AdamW optimizer~\cite{loshchilov2018decoupled} using fp16 precision for 30 epochs.

\begin{table}[]
    \centering \small
    \begin{tabular}{@{}ccccc@{}}
    \toprule
    \multicolumn{1}{l}{} & \multicolumn{1}{l}{} & \textbf{Asking-for-help} & \textbf{Others} & \textbf{Macro Avg} \\ \cmidrule{2-5}
    \multicolumn{1}{l}{} & \textbf{Support} & 22 & 376 & 298 \\ \midrule 
    \multirow{3}{*}{\textbf{\begin{tabular}[c]{@{}c@{}}Threshold\\ = 0.5\end{tabular}}} & \textbf{Precision} & 0.52 & 0.97 & 0.75 \\
     & \textbf{Recall} & 0.55 & 0.97 & 0.76 \\
     & \textbf{F1} & 0.53 & 0.97 & 0.75 \\ \midrule
    \multirow{3}{*}{\textbf{\begin{tabular}[c]{@{}c@{}}Threshold\\ = 0.0007\end{tabular}}} & \textbf{Precision} & 0.35 & 0.99 & 0.67 \\
     & \textbf{Recall} & 0.77 & 0.91 & 0.84 \\
     & \textbf{F1} & 0.48 & 0.95 & 0.71 \\ \bottomrule
    \end{tabular}
    \caption{Asking-for-help classification performance on the validation set. We searched for a decision threshold in which Asking-for-help recall was higher than 0.7 to encourage the Asking-for-help coverage rate.}
    \label{tab:classification-result}
\end{table}




\paragraph{Evaluating the Classifier.}
We evaluated the model every 50 steps and kept the checkpoint with the highest macro f1-score.
\Cref{tab:classification-result} shows the classification performance on the validation set.
To avoid unintentionally limiting question types, we adjusted the decision threshold (0.0007) to increase \texttt{Asking-for-help} recall to 0.7.
The decision threshold was decided by moving the decision threshold from 0.5 to 0 (we moved 5e-5 every step, \eg, 0.5, 0.49995, 0.49990, $\cdots$) and computed \texttt{Asking-for-help} recall.
We stopped the process and kept the decision threshold once \texttt{Asking-for-help} recall reached 0.7 ($\geq 0.7$).
The classification performance using 0.0007 as the threshold is shown in \Cref{tab:classification-result}.

\paragraph{Extracting Asking-for-Help Questions.}
We applied this text classifier on the entire \texttt{one-million-reddit-questions} dataset. It identified 129,483 asking-for-help questions (12.94\% of the entire Reddit dataset).


\begin{table*}
\small
\centering
\begin{tabular}{@{}lll@{}}
\toprule
\textbf{\#} & \textbf{Aspect} & \textbf{Survey Question} \\ \midrule
Q1 & \qone & \begin{tabular}[c]{@{}l@{}}If I have this question, I would reach out to other people, such as friends, family members, colleagues,\\ or experts, to get help. As compared to finding the answers by looking up information by myself with\\ a computer or a smartphone.\\(1) Very Unlikely  (2) Unlikely 
 (3) Neutral  (4) Likely  (5) Very Likely \end{tabular} \\ \midrule
Q2 & \qtwo & \begin{tabular}[c]{@{}l@{}}This question is asking for a response within a limited category. For example, questions in the following\\ categories: Yes/No, ``Does anybody else...'', Either/or, ``Would you rather...'', polls, surveys and\\ fill-in-the-blank questions.\\(1) Strongly Disagree 
 (2) Disagree  (3) Neutral  (4) Agree  (5) Strongly Agree\end{tabular} \\ \midrule
Q3 & \qthree & \begin{tabular}[c]{@{}l@{}}This question is more about encouraging responders to express their diverse experiences, opinions, or \\ preferences than actually getting help.\\(1) Strongly Disagree  (2) Disagree  (3) Neutral  (4) Agree  (5) Strongly Agree\end{tabular} \\ \midrule
Q4 & \qfour & \begin{tabular}[c]{@{}l@{}}This question requires or encourages the responders to further discuss with the asker in order to come up\\ with an appropriate answer.\\(1) Strongly Disagree 
 (2) Disagree  (3) Neutral  (4) Agree  (5) Strongly Agree\end{tabular} \\ \midrule
Q5 & \qfive & \begin{tabular}[c]{@{}l@{}}Without reaching out to other people, for a layperson with no background knowledge related to this question.\\ How long do you think it would likely take for them to figure out the answer to this question with access to\\ the internet?\\(1) $\leq$ 30 minutes (2) 30 minutes-2 hours (3) 2 hours-half a day (4) half a day-1 day (5) $\geq$ 1 day (6) Undoable\end{tabular} \\ \midrule
Q6 & \qsix & \begin{tabular}[c]{@{}l@{}}If I have this question, I would prefer to have a conversation regarding the details of the question and\\ have a further discussion with the answerer. As compared to asking the question as is and waiting for\\ the answers.\\(1) Strongly Disagree  (2) Disagree  (3) Neutral  (4) Agree  (5) Strongly Agree\avon{discussed with Alan.}\avon{"As compared..." I can't understand this sentence. Is it "As compared to asking the question and waiting for the answers?"}\alan{yes}\end{tabular} \\ \midrule
Q7 & \qseven & \begin{tabular}[c]{@{}l@{}}Suppose I asked this question using a computer or smartphone instead of making phone calls or in-person\\ sessions. In that case, I prefer the answer to be provided through a conversation \eg, via WhatsApp or\\ other messaging applications) compared to other written formats, such as emails, social media replies, or\\ online forums.\\(1) Strongly Disagree  (2) Disagree 
 (3) Neutral  (4) Agree  (5) Strongly Agree\end{tabular} \\ \midrule
Q8 & \qeight & \begin{tabular}[c]{@{}l@{}}Without reaching out to other people to get help, I will be able to answer the question by looking up\\ information by myself with access to a computer or a smartphone.\\(1) Very Difficult  (2) Difficult  (3) Neutral  (4) Easy  (5) Very Easy\end{tabular} \\ \bottomrule
\end{tabular}
\caption{The eight categories and the inquiries used to collect workers' opinions.}
\label{tab:8-questions}
\end{table*}

\subsection{Collecting Human Opinions About Questions}
From the 129,483 asking-for-help questions, 500 questions were randomly sampled for human annotation on Amazon Mechanical Turk (MTurk).
Five hundred out of one million questions calculated from a 95\% confidence level and a 5\% margin of error provided a valid sample size for analysis~\cite{israel1992determining}.\cy{Do we have citations about how this is estimated?}
For each of the 500 questions, we asked nine workers to rate eight aspects using a five-point Likert Scale ranging from (1) Strongly Disagree to (5) Strongly Agree.
Table~\ref{tab:8-questions} shows the eight aspects we used.
Options for Q1 ranged from (1) Very Unlikely to (5) Very Likely.
Options for Q5 were 
(1) 30 minutes or less, 
(2) 30 minutes-2 hours, 
(3) 2 hours-half a day, 
(4) half a day-1 day, 
(5) 1 day or more, and 
(6) Undoable.
Options for Q8 ranged from (1) Very Difficult to (5) Very Easy.
\avon{discussed with Alan}\avon{options for Q5 might be able to integrate into table 4?}

In the study, each Human Intelligence Task (HIT) contained one asking-for-help question for which each worker was asked to answer the eight survey questions (Table~\ref{tab:8-questions}).
\Cref{fig:annotation-interface} (see Appendix) shows the worker interface.
We added a 90-second submission lock on the interface to prevent malicious workers from spamming.
The compensation for one HIT assignment was \$0.25, which was estimated using an hourly wage of \$10.
Four built-in MTurk qualifications were also used: Locale (US Only), HIT Approval Rate ($\geq$98\%), Number of Approved HITs ($\geq$3000), and Adult Content Qualification.

\subsection{Categorizing Asking-For-Help Questions}

\begin{table*}
    \centering \small
    \begin{tabular}{llrl}
        \toprule
        \multicolumn{1}{c}{\textbf{Rank}} & \multicolumn{1}{c}{\textbf{Category}} & \multicolumn{1}{c}{\textbf{\# Questions}} & \multicolumn{1}{c}{\textbf{Brief Description}} \\ \midrule
        1 & tech                              & 67  &Technology\\
        2 & reddit\_tech                      & 47  &Reddit-related, Reddit searching, Reddit tech support  \\
        3 & medical\_health\_diet             & 40  &Medical, health, or diet\\
        4 & dailylife\_hack\_forfun\_home     & 36  &Daily life, home, life hack, for fun or ``food for thought'' discussion \\
        5 & movie\_music\_media\_hobby\_sport & 31  &Movies, music, media, hobby, sport \\
        6 & help\_find\_things                & 28  &Help find things or information by providing description  \\
        7 & culture\_believe\_language        & 20  &Culture, beliefs, language  \\
        8 & socializing                       & 17  &Socializing  \\
        9 & country\_location\_travel         & 17  &Country, location, traveling  \\
        10 & life\_suggestion                  & 16  &General life suggestions \\
        
        11 & science                           & 13  & Science  \\
        12 & history\_old\_days\_future        & 12  & History and discussion about the past or future \\
        13 & career                            & 11  & Career  \\
        14 & food                              & 10  & Food \\
        15 & human\_body                        & 10  & Human body mechanism and functions\\
        16 & legal\_regulation                  & 10  & Law, legal questions, general regulations\\
        17 & news\_events                       & 9   & News and real-life events\\
        18 & social\_etiquette                  & 9   & Social etiquette\\
        19 & mental\_health                     & 8   & Mental health\\
        20 & learning\_skills                   & 7   & Learning and acquire skills\\
        21 & personal\_finance                  & 7   & Personal finance\\
        22 & worldwide\_society\_effect\_impact & 7   & Worldwide scale discussion/societal changes and impact\\
        23 & NSFW\_sensitive                    & 6   & Not safe for work; not suitable for work and sensitive topics\\
        24 & politics                          & 6   & Politics\\
        25 & relationship                      & 6   & Couple relationships\\
        26 & religion                          & 5   & Religion\\
        - & Others                            & 45  & Categories that appear fewer than five times \\
        \bottomrule
    \end{tabular}
    \caption{
        The frequency of the coded categories. Categories with less than five questions are merged into the ``Other'' category, resulting in a total of 27 categories.
    }
    \label{tab:category}
\end{table*}

One author (A1) went through all the 500 sampled questions to get familiar with the data. 
Open coding was performed to come up with a coding scheme. 
The process was performed repeatedly until all questions are categorized. A total of 44 mutually exclusive categories were created, and each question belonged to only one category. For simplicity, we merged categories that contain less than five questions into the ``Other'' category, resulting in a total of 27 categories. 
\Cref{tab:category} shows the frequency of the coded categories.
Following the coding scheme, another author (A2) independently coded 100 randomly sampled questions from the 500 asking-for-help questions. The inter-coder reliability reached a Cohen's kappa of 0.574.

\section{Experimental Results}

\begin{figure*}
    \centering
    \includegraphics[width=0.999\linewidth]{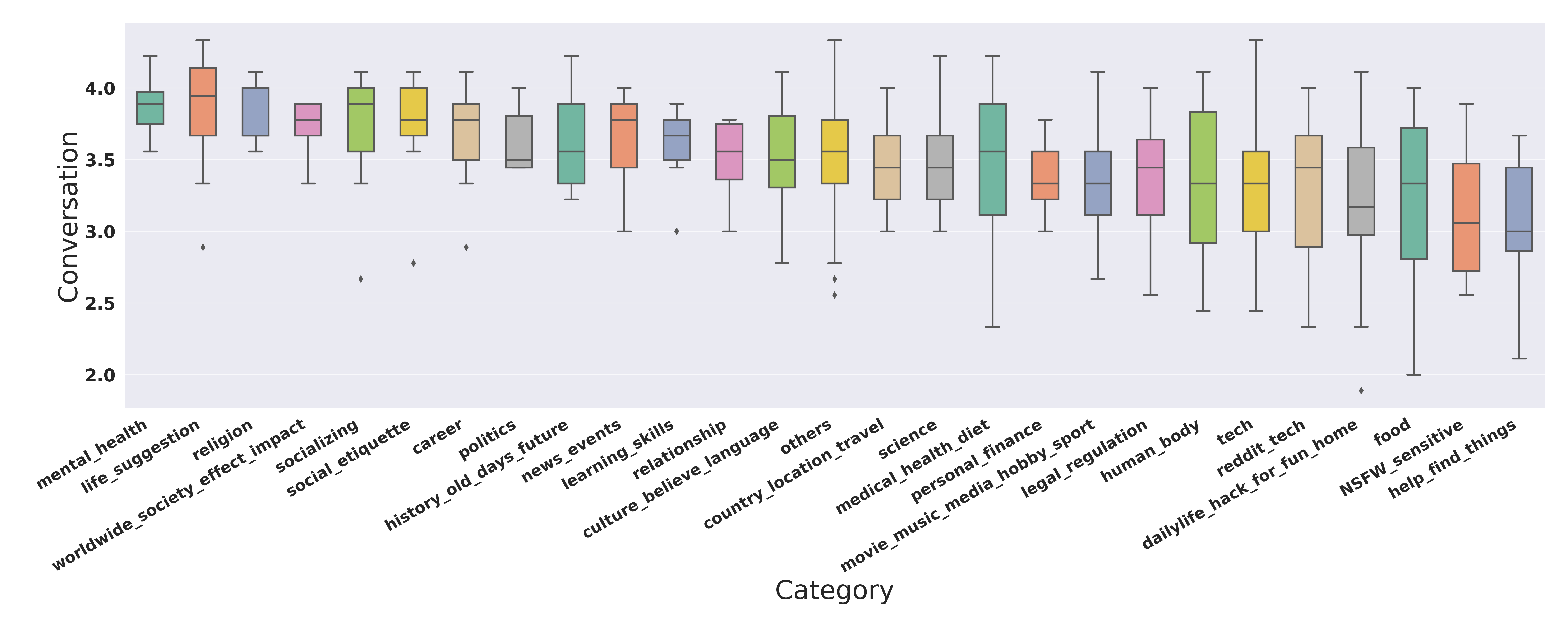}
    \caption{Conversation score (Q6) distribution over different categories. People like to have conversations to consult on questions that do not have clear answers (\eg, mental\_health and life\_suggestion). For questions that have clearer answers (\eg, help\_find\_things and tech), conversation is less needed.}
    \Description{Box plot of Conversation score (Q6) distribution over different categories. The Y-axis value is the score, X-axis are the box plot different categories. From left to right, category and values of the box plot are:
    mental\_health with minimum value of 3.56, first quartile of 3.75, median of 3.89, third quartile of 3.97, and maximum value of 4.22, 
    life\_suggestion with minimum value of 2.89, first quartile of 3.67, median of 3.94, third quartile of 4.14, and maximum value of 4.33, 
    religion with minimum value of 3.56, first quartile of 3.67, median of 3.67, third quartile of 4.0, and maximum value of 4.11, 
    worldwide\_society\_effect\_impact with minimum value of 3.33, first quartile of 3.67, median of 3.78, third quartile of 3.89, and maximum value of 3.89, 
    socializing with minimum value of 2.67, first quartile of 3.56, median of 3.89, third quartile of 4.0, and maximum value of 4.11, 
    social\_etiquette with minimum value of 2.78, first quartile of 3.67, median of 3.78, third quartile of 4.0, and maximum value of 4.11, 
    career with minimum value of 2.89, first quartile of 3.5, median of 3.78, third quartile of 3.89, and maximum value of 4.11, 
    politics with minimum value of 3.44, first quartile of 3.44, median of 3.5, third quartile of 3.81, and maximum value of 4.0, 
    history\_old\_days\_future with minimum value of 3.22, first quartile of 3.33, median of  3.56, third quartile of 3.89, and maximum value of 4.22, 
    news\_events with minimum value of 3.0, first quartile of 3.44, median of 3.78, third quartile of 3.89, and maximum value of 4.0, 
    learning\_skills with minimum value of 3.0, first quartile of 3.5, median of 3.67, third quartile of 3.78, and maximum value of 3.89, 
    relationship with minimum value of 3.0, first quartile of 3.36, median of 3.56, third quartile of 3.75, and maximum value of 3.78, 
    culture\_believe\_language with minimum value of 2.78, first quartile of 3.31, median of 3.5, third quartile of 3.81, and maximum value of 4.11, 
    others with with minimum value of 2.56, first quartile of 3.33, median of 3.56, third quartile of 3.78, and maximum value of 4.33, 
    country\_location\_travel with minimum value of 3.0, first quartile of 3.22, median of 3.44, third quartile of 3.67, and maximum value of 4.0, 
    science with minimum value of 3.0, first quartile of 3.22, median of 3.44, third quartile of 3.67, and maximum value of 4.22, 
    medical\_health\_diet with minimum value of 2.33, first quartile of 3.11, median of 3.56, third quartile of 3.89, and maximum value of 4.22, 
    personal\_finance with minimum value of 3.0, first quartile of 3.22, median of 3.33, third quartile of 3.56, and maximum value of 3.78, 
    movie\_music\_media\_hobby\_sport with minimum value of 2.67, first quartile of 3.11, median of 3.33, third quartile of 3.56, and maximum value of 4.11, 
    legal\_regulation with minimum value of 2.56, first quartile of 3.11, median of 3.44, third quartile of 3.64, and maximum value of 4.0, 
    human\_body with minimum value of 2.44, first quartile of 2.92, median of 3.33, third quartile of 3.83, and maximum value of 4.11, 
    tech with minimum value of 2.44, first quartile of 3.0, median of 3.33, third quartile of 3.56, and maximum value of 4.33, 
    reddit\_tech with minimum value of 2.33, first quartile of 2.89, median of 3.44, third quartile of 3.67, and maximum value of 4.0, 
    dailylife\_hack\_for\_fun\_home with minimum value of 1.89, first quartile of 2.97, median of 3.17, third quartile of 3.58, and maximum value of 4.11, 
    food with minimum value of 2.0, first quartile of 2.81, median of 3.33, third quartile of 3.72, and maximum value of 4.0, 
    NSFW\_sensitive with minimum value of 2.56, first quartile of 2.72, median of 3.06, third quartile of  3.47, and maximum value of 3.89, 
    help\_find\_things with minimum value of 2.11, first quartile of 2.86, median of 3.0, third quartile of 3.44, and maximum value of 3.67.}
    \label{fig:score-q6-conversation}
\end{figure*}
By comparing the annotated aspects and categories, we formulated three results.

\begin{table*}[h]
    \centering \small
    \begin{tabular}{lp{11cm}}
        \toprule
        \multicolumn{1}{c}{\textbf{Category}} & \multicolumn{1}{c}{\textbf{Question}} \\ \midrule
        tech                                  &Recommendation for a vacuum cleaner?  \\ \midrule
        life\_suggestion                      &What to do when you are feeling lost in life?  \\ \midrule
        work\_place                            &What is the best way to subtly and consistently annoy your coworkers, without them ever realising it's your fault?  \\ \midrule
        others                                &What happens after we die ?  \\ \midrule
        mental\_health                        &What do you do when you can’t get an anxiety-inducing thought out of your head? \\ \midrule
        medical\_health\_diet                 &What happens when you chew a poisonous flower for a few seconds but spit it out? \\ \midrule
        medical\_health\_diet	              &how much pain did you feel after wisdom tooth removal? \\ \midrule
        medical\_health\_diet                 &How to reduce one sided cheek fat? \\ \midrule
        life\_suggestion                      &How to control emotions? How do people control their emotions when they lost their loved one? \\ \midrule
        life\_suggestion                       &What steps should I take towards moving out of my parents house? I'm at the ripe old age of 16 when the state of Pennsylvania graciously gives me the chance to operate a motor vehicle. What can I do to get myself headed in the direction of living on my own?  \\ \midrule
        life\_suggestion                      &Is it possible to make a good situation out of any bad situation?  \\ \midrule
        science                               &What is a fine tuned universe? Why is gravity fine-tuned? \\ \midrule
        history\_old\_days\_future             &Is politics more entertaining now than it was in decades prior?  \\ \midrule
        mental\_health                        &Let me start off by saying, yes I've tried most of the normal avenues, and yet my mind is still filled with thoughts of nihilism. Every moment of my life feels like I'm just waiting. Not for anything in particular, just something. Is there anywhere for people like me to go, and just disappear?  \\ \midrule
        others                               &This housing market is wild. Is it going to last the next 4 years? \\ \bottomrule
        \end{tabular}
    \caption{Questions with the highest Conversation (Q6) score ($\geq 4.22$).}
    \label{tab:top_examples}
\end{table*}


\subsection{What types of questions are a better fit for conversational UI?}
\Cref{fig:score-q6-conversation} shows the box chart of the Conversation scores \alan{(based on five-point Likert Scale)} over different categories.
The categories were sorted descending (from left to right) using the mean Conversation scores.
We found that people believe conversations were needed most\avon{x}\avon{Conversation score >=?} when questions did not have clear answers, \eg, 
mental\_health, life\_suggestion, religion, worldwide\_society\_effect\_impact, socializing, and social\_etiquette.
Questions that might have concrete responses did not need to be resolved through conversations\avon{x}\avon{Conversation score >=?}, \eg, 
help\_find\_things, NSFW\_sensitive, food, dailylife\_hack\_for\_fun\_home, and reddit\_tech.
See \Cref{tab:top_examples} for questions with the highest Conversation score.

\begin{table*}
    \centering \small
    \begin{tabular}{
        lccccccc
    }
    \toprule
    & \textbf{\qtwo} & \textbf{\qthree} & \textbf{\qfour} & \textbf{\qfive} & \textbf{\qsix} & \textbf{\qseven} & \textbf{\qeight} \\ \midrule
    \textbf{\qone} & 0.038 & \,\bfseries 0.519 & \,\bfseries 0.553 & \,\,0.191 & \,\,0.469 & \,\bfseries 0.502  & -0.035 \\
    \textbf{\qtwo} &\multicolumn{1}{c}{-}& -0.103 & -0.089 & -0.104 & -0.111 & -0.016 & \,\,0.321 \\
    \textbf{\qthree} &\multicolumn{1}{c}{-}&\multicolumn{1}{c}{-}& \,\bfseries 0.658 & \,\,0.246 & \,\bfseries 0.663 & \,\bfseries 0.607 & -0.186 \\
    \textbf{\qfour} &\multicolumn{1}{c}{-}&\multicolumn{1}{c}{-}&\multicolumn{1}{c}{-}& \,\,0.323 & \,\bfseries 0.720 & \,\bfseries 0.672 & -0.326 \\
    \textbf{\qfive} &\multicolumn{1}{c}{-}&\multicolumn{1}{c}{-}&\multicolumn{1}{c}{-}&\multicolumn{1}{c}{-}& \,\,0.294 & \,\,0.259 & -0.278 \\
    \textbf{\qsix} &\multicolumn{1}{c}{-}&\multicolumn{1}{c}{-}&\multicolumn{1}{c}{-}&\multicolumn{1}{c}{-}&\multicolumn{1}{c}{-}& \,\bfseries 0.665 & -0.323 \\
    \textbf{\qseven} &\multicolumn{1}{c}{-}&\multicolumn{1}{c}{-}&\multicolumn{1}{c}{-}&\multicolumn{1}{c}{-}&\multicolumn{1}{c}{-}&\multicolumn{1}{c}{-}& -0.224 \\ \bottomrule
    \end{tabular}
    \caption{Pearson Correlation between different aspects. \textbf{Bold} represents highly correlated ($\geq 0.5$).}
    \label{tab:correlation}
\end{table*}




\subsection{Correlation among aspects}
To see the relationships among different aspects, we computed the Pearson correlation between all the aspects.
\Cref{tab:correlation} shows the correlation. 
We found that \qsix is highly correlated with \qthree (0.663), \qfour (0.720), and \qseven (0.665), 
suggesting that when a question required a conversation to satisfactorily explore,
people believe this question to 
(\textit{i}) be more related to personal opinions and experiences and 
(\textit{ii}) require more discussion.
Also, in such cases people generally prefer to have a conversation on messaging applications compared to other formats.
\avon{discussed with Alan}%
\avon{In the method section, you said "Questions that were not considered as asking for help questions were those that are based on the personal opinions and experiences of the answerer." But it just showed up here. Maybe use another phrasing to describe it in the method section?}%
Since the score for Difficulty is in reverse fashion, (1) being Very Difficult and (5) being Very Easy, the Difficulty score is negatively correlated with most other aspects.


\begin{figure*}
    \centering
    \includegraphics[width=0.999\linewidth]{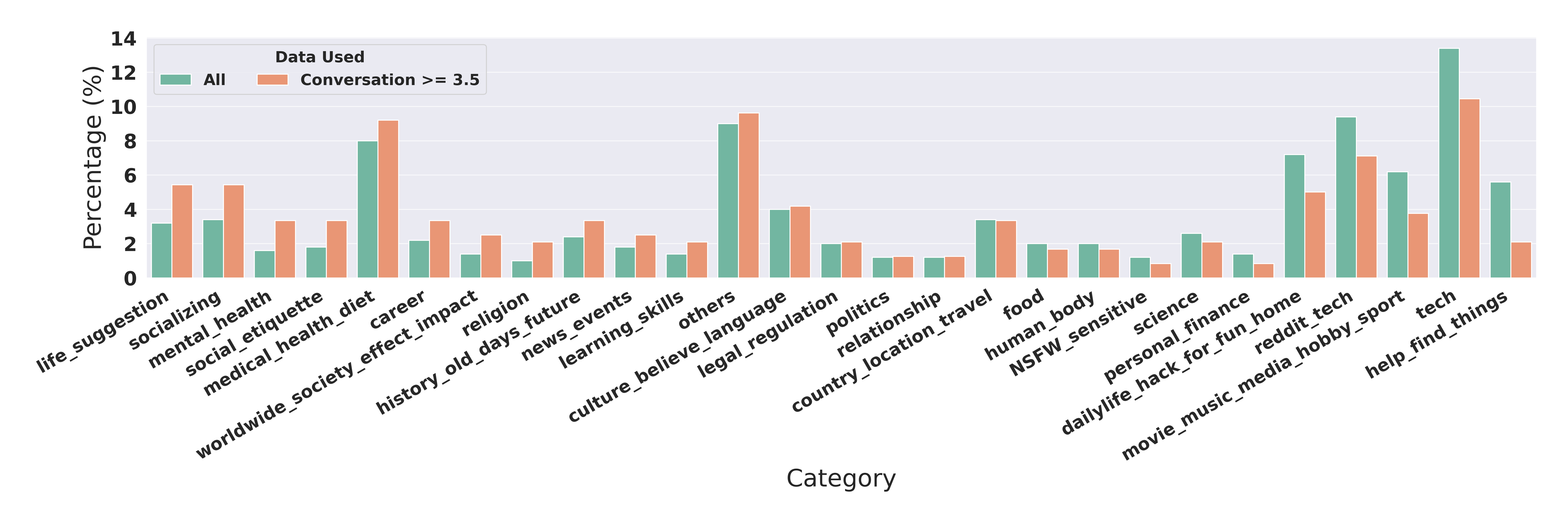}
    \caption{Category distribution shift between all the questions and the ones with higher conversation preference (Conversation score (Q6) $\geq 3.5$). Categories with the highest increase in percentage are life, mental health, and socializing.}
    \Description{Bar chart of distribution shift between all questions and ones with Q6 score larger than 3.5. The Y-axis value is the percentage. The X-axis are two bars for each of the different categories, representing percentage within all questions (presented in a green bar) and percentage within Q6 score larger than 3.5 (presented in an orange bar). From left to right, category and values of the two bars are: 
    life\_suggestion with green bar value of 3.2 and orange bar value of 5.44, 
    socializing with green bar value of 3.4 and orange bar value of 5.44, 
    mental\_health with green bar value of 1.6 and orange bar value of 3.35, 
    social\_etiquette with green bar value of 1.8 and orange bar value of 3.35, 
    medical\_health\_diet with green bar value of 8 and orange bar value of 9.21, 
    career with green bar value of 2.2 and orange bar value of 3.35, 
    worldwide\_society\_effect\_impact with green bar value of 1.4 and orange bar value of 2.51, 
    religion with green bar value of 1 and orange bar value of 2.09, 
    history\_old\_days\_future with green bar value of 2.4 and orange bar value of 3.35, 
    news\_events with green bar value of 1.8 and orange bar value of 2.51, 
    learning\_skills with green bar value of 1.4 and orange bar value of 2.09, 
    others with green bar value of 9 and orange bar value of 9.62, 
    culture\_believe\_language with green bar value of 4 and orange bar value of 4.18, 
    legal\_regulation with green bar value of 2 and orange bar value of 2.09, 
    politics with green bar value of 1.2 and orange bar value of 1.26, 
    relationship with green bar value of 1.2 and orange bar value of 1.26, 
    country\_location\_travel with green bar value of 3.4 and orange bar value of 3.35, 
    food with green bar value of 2 and orange bar value of 1.67, 
    human\_body with green bar value of 2 and orange bar value of 1.67, 
    NSFW\_sensitive with green bar value of 1.2 and orange bar value of 0.84, 
    science with green bar value of 2.6 and orange bar value of 2.09, 
    personal\_finance with green bar value of 1.4 and orange bar value of 0.84, 
    dailylife\_hack\_for\_fun\_home with green bar value of 7.2 and orange bar value of 5.02, 
    reddit\_tech with green bar value of 9.4 and orange bar value of 7.11, 
    movie\_music\_media\_hobby\_sport with green bar value of 6.2 and orange bar value of 3.77, 
    tech with green bar value of 13.4 and orange bar value of 10.46, 
    help\_find\_things with green bar value of 5.6 and orange bar value of 2.09.
    }
    \label{fig:distribution-shift}
\end{figure*}


\subsection{Category distribution shift}
We further compared how the categories were distributed among all the questions and among the questions with high Conversation scores.
The category distribution was represented by the percentage of questions within different categories.
Questions with high conversation preference (Conversation-desiring) were determined by Conversation score $\geq 3.5$. 
\Cref{fig:distribution-shift} shows the distribution shift.
We sorted the categories by the difference between the distribution shifts (\ie, percentage of Conversation-desiring question subtracting percentage in All question) descending from left to right. 
The figure suggests that the life\_suggestion, socializing, and mental\_health categories increased more within the Conversation-desiring questions
while movie\_music\_media\_hobby\_sport, tech, and help\_find\_things reduced more.
This further implied that more personal or social questions are better suited for conversation compared to other types.






\section{Discussion}

From our results, we identified three areas of discussion.
\paragraph{People want to talk about social situations and personal problems.}
Our analysis shows that questions people believe require conversation to resolve satisfactorily are \textbf{highly social and personal}.
Examples include life suggestions, socializing, and mental health.
Meanwhile, the questions related to tech or information seeking were considered least requiring of conversation. 

These findings prompt us to rethink the notion of conversation, especially the differences between producing answers in fluent, natural language and exploring a topic in back-and-forth interaction.
Under the broader umbrella of \convsys, many techniques were created and evolved to achieve the latter, but our findings suggested that in some cases, users might only need the former.
Our second conclusion is that
the possibility of enabling multiple response channels for \convsys could be further pursued.
For the questions that tend not to require an interactive conversation to resolve, the system can take extra time and resources to prepare the answer and respond via an alternative channel such as email or text messages.
This will introduce a new set of technical and UX questions, including how to automatically choose the response channel, customize the user's preference, collect needed information, or ask follow-up questions via multiple channels.
Finally, we are aware that a significant body of work has explored using chatbots or conversational agents to provide therapy or mental health support~\cite{fitzpatrick2017delivering,10.1145/3392836}\cy{If there are a lot of works, we can probably cite some more papers.}\kenneth{I added one more.}.
Even though these questions are often much harder to solve, our results suggest that these attempts are valuable to users.

\paragraph{Some questions require extra attention.}
Some topics are sensitive, controversial, or potentially harmful.
Categories such as politics, religion, NSFW\_sensitive, and suicide-related likely need to be handled with extra caution.
Our study showed that these types of questions are not rare.
Out of 500 asking-for-help questions, six were about politics, five were about religion, six were categorized as NSFW, and one was related to suicide.


\paragraph{Limitations.}
We are aware of some limitations of this work.
First, the automatic classifiers' performance was not perfect.
Although we tuned the classifier's parameter to yield high recall, some asking-for-help questions may have been excluded from our study.
Second, the scale of our system was relatively small.
We could only afford to manually annotate and categorize 500 (each question having nine responses) of the millions of questions posted to AskReddit.
Finally, the selection of platforms inevitably imposed biases.
The asking-for-help questions were sampled from Reddit, whose users tend to be younger, US-centric, and primarily male~\cite{reddit-demographic}.
The AskReddit platform also has community norms that encourage questions that generate discussions rather than asking-for-help questions.
Using MTurk introduced similar biases.

\section{Conclusion and Future Work}

This paper studies what types of questions are most suitable for conversational modality.
We recruited online crowd workers to answer eight inquiries about 500 questions posted on AskReddit and performed an in-depth analysis.
We found that the questions people believe require conversation to resolve satisfactorily are highly social and personal.
Examples include life suggestions, socializing, and mental health.
Meanwhile, the questions related to tech or information seeking were considered least requiring of conversation.
In the future, we will develop computational models that automatically recommend the appropriate delivery modality for questions. 
Such a model would allow intelligent question-answering systems to personalize the communication channel to users.

\begin{acks}
We are grateful to the anonymous reviewers for their constructive feedback, and to the MTurk workers for their participation in our study.
\end{acks}

\bibliographystyle{ACM-Reference-Format}
\bibliography{bib/custom,bib/sample-base,bib/software}

\appendix
\section{Supplementary Material}
\begin{figure*}
    \centering
    \includegraphics[height=0.95\textheight]{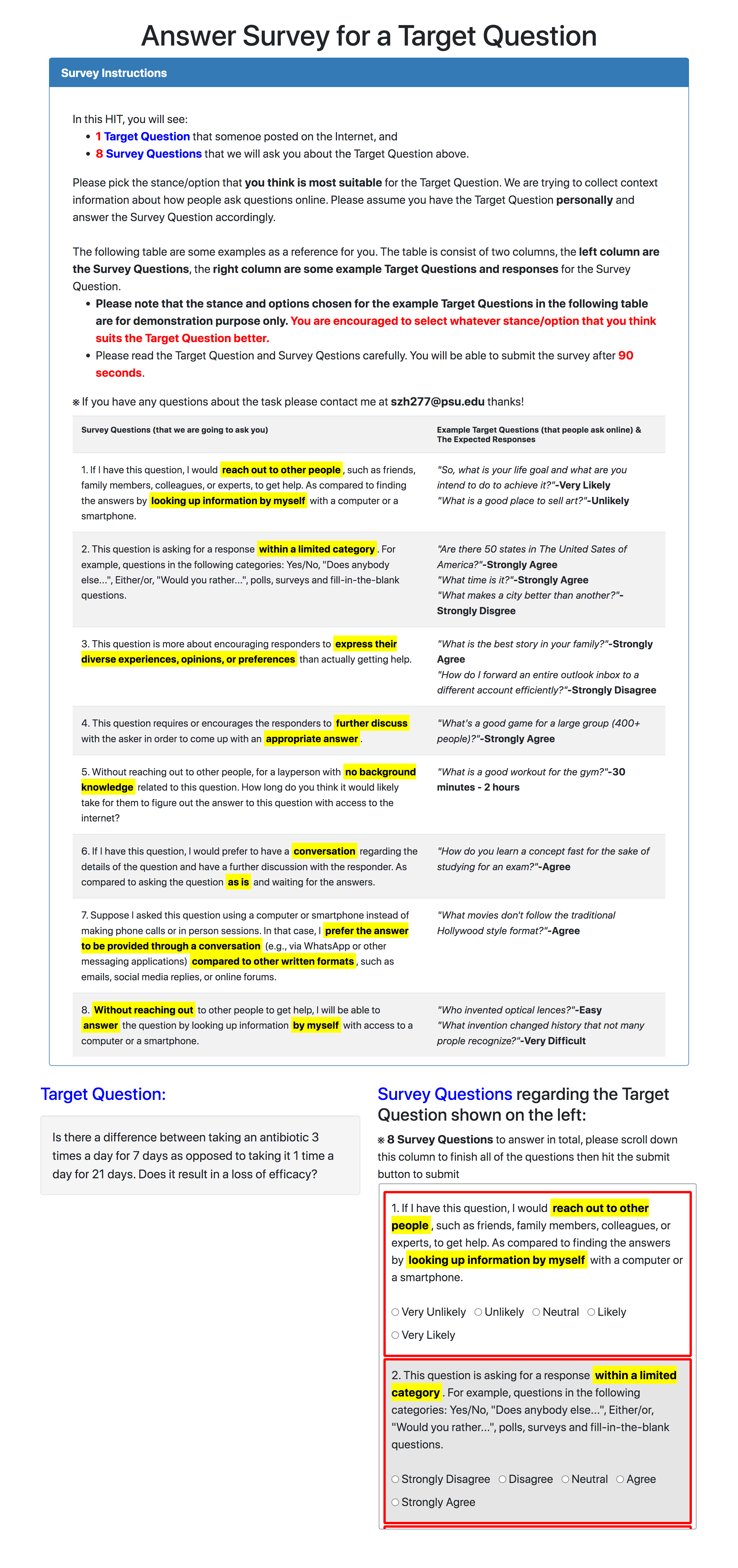}
    \caption{Interface for MTurk workers.}
    \Description{There are three separate sections. One one the top, one at the bottom left, and one at the bottom right. The top section provides instructions for the MTurk workers accompanied with examples for each of the 8 inquiries. The bottom left section shows the target question. The bottom right section are the 8 inquiries we ask MTurk workers to provide their opinions.}
    \label{fig:annotation-interface}
\end{figure*}

\Cref{fig:annotation-interface} shows the MTurk interface for collecting online crowd workers' opinions through the eight questions (\Cref{tab:8-questions}).

\end{document}